\documentclass[aps, prb, reprint, superscriptaddress, amsmath, amssymb]{revtex4-2}

\usepackage{graphicx}
\usepackage[T1]{fontenc}
\usepackage{newtx}
\usepackage{xcolor}
\usepackage[colorlinks, allcolors=blue]{hyperref}
\usepackage{orcidlink}

\begin{document}

\title{Lattice dynamics and the spectroscopic signatures of H-bond disorder in \texorpdfstring{$\delta$}{delta}-AlOOH}

\author{Chenxing Luo\,\orcidlink{0000-0003-4116-6851}}
\email{cl3658@columbia.edu}
\affiliation{Department of Applied Physics and Applied Mathematics, Columbia University, New York, New York 10027, USA}
\affiliation{Department of Geosciences, Princeton University, Princeton, New Jersey 08544, USA}

\author{Sangjoon Lee\,\orcidlink{0000-0002-2367-3932}}
\altaffiliation[Current address: ]{Department of Materials Science and Engineering, Stanford University, Stanford, California 94305, USA}
\affiliation{Department of Applied Physics and Applied Mathematics, Columbia University, New York, New York 10027, USA}

\author{Hongjin Wang\,\orcidlink{0009-0008-9766-1177}}
\affiliation{Department of Applied Physics and Applied Mathematics, Columbia University, New York, New York 10027, USA}

\author{Zhen Zhang\,\orcidlink{0009-0001-0810-8054}}
\affiliation{Department of Physics and Astronomy, Iowa State University, Ames, Iowa 50011, USA}

\author{Renata Wentzcovitch\,\orcidlink{0000-0001-5663-9426}}
\email{rmw2150@columbia.edu}
\affiliation{Department of Applied Physics and Applied Mathematics, Columbia University, New York, New York 10027, USA}
\affiliation{Department of Earth and Environmental Sciences, Columbia University, New York, New York 10027, USA}
\affiliation{Lamont--Doherty Earth Observatory, Columbia University, Palisades, New York 10964, USA}
\affiliation{Data Science Institute, Columbia University, New York, New York 10027, USA}

\date{\today}

\begin{abstract}
Compression of $\delta$-AlOOH produces a multistage evolution of H-bond configurations, average symmetry, and proton centering. Raman and infrared anomalies at 5--10~GPa mark early changes in phonon frequencies and linewidths within this sequence, while static harmonic phonons soften only at much higher pressure. We combine harmonic phonon calculations with strongly constrained and appropriately normed (SCAN)-based deep-potential molecular dynamics and phonon quasiparticle analysis at 300~K. Vibrational spectra show branch reorganization and linewidth growth during H-bond disordering. Loss of coherence in a coupled proton-lattice mode develops before the bulk compression anomaly and the change in average symmetry near 9--10~GPa; full proton centering occurs later, at 16--18~GPa. Quasiparticle renormalization and directional longitudinal-optical--transverse-optical (LO-TO) splitting reproduce the three higher-frequency components of the ambient-pressure OH-stretching Raman multiplet; the lowest-frequency component remains unresolved. Finite-temperature H-bond dynamics and loss of spectral coherence are necessary to explain the early spectroscopic anomalies.
\end{abstract}

\maketitle

\section{Introduction}

\begin{figure}[htbp]
    \centering
    \includegraphics[width=0.3\textwidth]{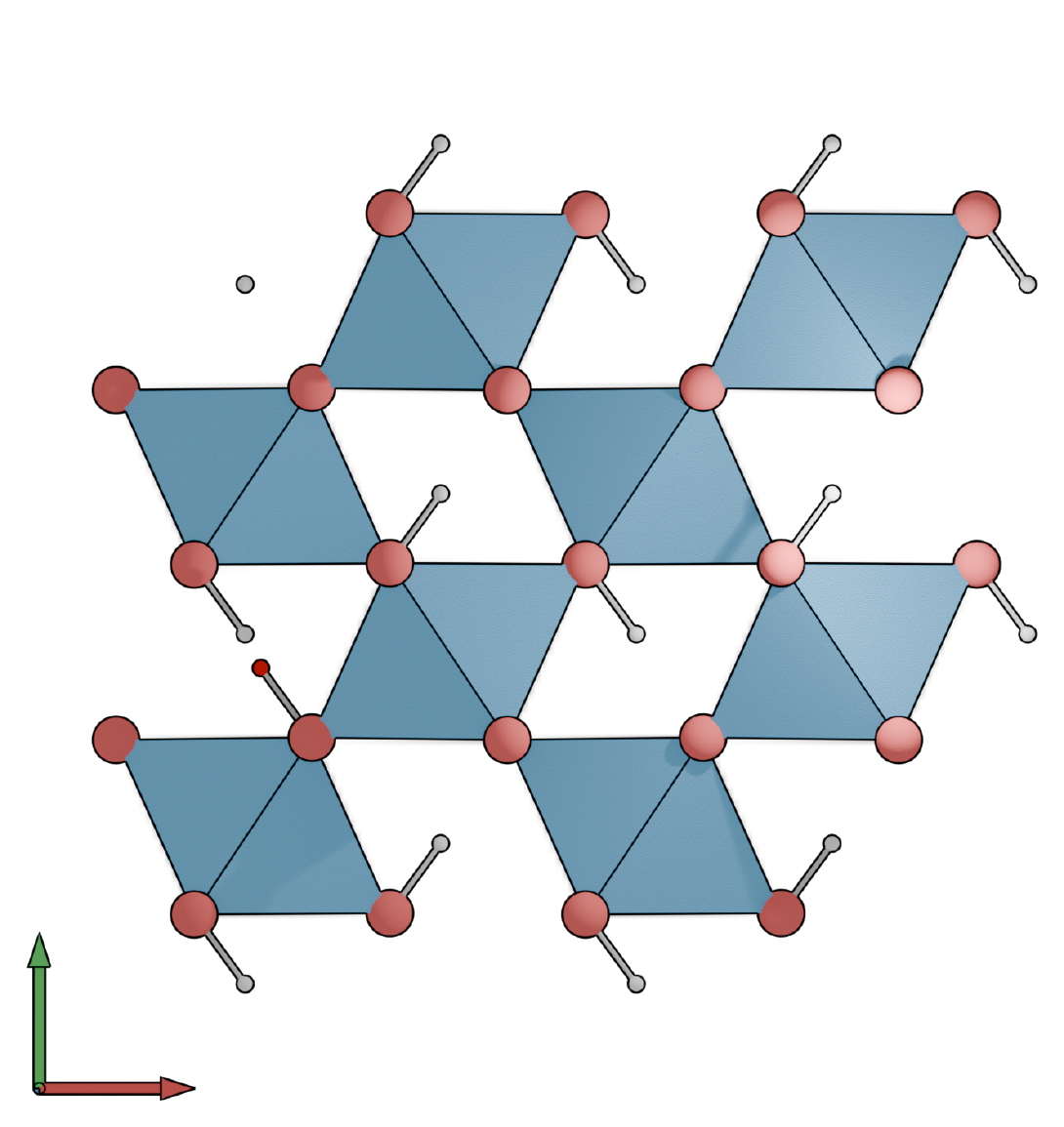}
    \caption{Top view of low-pressure $\delta$-AlOOH in the $\mathrm{P}2_1\mathrm{nm}$ structure. $\mathrm{AlO}_6$ octahedra are shown as polyhedra, O atoms as red spheres, and H atoms as small gray spheres. Red H markers indicate the symmetry-related off-center proton sites associated with H-bond disordering.}
    \label{fig:disorder}
\end{figure}

$\delta$-AlOOH \cite{suzukiNewHydrousPhase2000, ohtaniStabilityFieldNew2001} ($\delta$) is a dense hydrous phase used to study hydrogen bonding under lower mantle conditions ($\sim$20--135~GPa; $\sim$2,000--4,000~K). At 300~K, compression produces a multistage evolution of H-bond configurations and proton positions, accompanied by anomalies in compression and vibrations \cite{tsuchiyaFirstPrinciplesCalculation2002, tsuchiyaVibrationalPropertiesDAlOOH2008, luoInitioInvestigationHbond2022, sano-furukawaDirectObservationSymmetrization2018}. Neutron diffraction resolves the evolution of the average proton distribution. At $\sim$9~GPa, the two off-center sites acquire equivalent occupations and average symmetry changes from $\mathrm{P}2_1\mathrm{nm}$ to $\mathrm{Pnnm}$ (Fig.~\ref{fig:disorder}). The distribution remains bimodal above this pressure and becomes centered only at $\sim$16--18~GPa \cite{sano-furukawaDirectObservationSymmetrization2018}. Similar changes in H-bond geometry and dynamics occur in other dense hydrous phases and high-pressure ice \cite{mookherjeeAnomalousElasticBehavior2019, tsuchiyaCrystalStructureEquation2015, benoitShapesProtonsHydrogen2005, benoitReassigningHydrogenBondCentering2002, zhangDeepNeuralNetwork2020, komatsuHydrogenBondSymmetrisation2024, cherubiniQuantumEffectsHbond2024}.

Raman and infrared (IR) spectra reveal how phonon frequencies and linewidths change during H-bond disordering and proton centering. Broad OH-stretching peaks are present at ambient pressure \cite{ohtaniStabilityFieldNew2001}. Under compression, anomalous pressure dependences, linewidth broadening, and the softening or disappearance of specific branches begin at $\sim$5--9~GPa \cite{kagiInfraredAbsorptionSpectra2010, mashinoSoundVelocitiesDAlOOH2016, wangRamanScatteringCr32022}, close to the disordering transition. Raman and luminescence measurements show only subtle changes at the later centering pressure of $\sim$16--18~GPa \cite{wangRamanScatteringCr32022}. Interpreting this separation between disordering and centering requires information about local motion beyond average symmetry. In compressed ice, spatial disorder among off-center sites, thermal hopping, quantum tunneling, and proton centering differ in local proton distributions and dynamics, even when they share an average space group \cite{komatsuHydrogenBondSymmetrisation2024}. Raman and IR spectra complement structural averages by probing vibrational restoring forces and coherence. We examine how frequency shifts and linewidths relate to configurational populations, bulk compressibility, and the average proton distribution in $\delta$-AlOOH.

Static harmonic theory misses these early frequency shifts and linewidth changes. Density functional theory (DFT) calculations using the Perdew--Burke--Ernzerhof generalized-gradient approximation (PBE-GGA) predict softening associated with full H-bond symmetrization at $\sim$30--40~GPa \cite{tsuchiyaElasticPropertiesDAlOOH2009, tsuchiyaFirstPrinciplesCalculation2002, tsuchiyaVibrationalPropertiesDAlOOH2008}. These calculations describe the loss of harmonic stability of an ordered structure on approach to proton centering. The experimental frequencies and damping change during H-bond disordering, while the ordered structures remain harmonically stable.

Finite-temperature atomic motion can change frequencies and damping before ordered harmonic modes soften. An earlier \textit{ab initio} molecular dynamics (MD) study including nuclear quantum effects through a quantum thermal bath found softening and fading of OH-stretching peaks at $\sim$10~GPa \cite{bronsteinThermalNuclearQuantum2017}. These results support a finite-temperature dynamical description, but leave the underlying branch-specific frequency shifts and linewidths unresolved.

Our multiconfigurational quasiharmonic study described disordering through the coexistence and pressure-dependent populations of H-bond arrangements \cite{luoInitioInvestigationHbond2022}. Arrangements with different proton alignments along [001] have distinct vibrational spectra and free energies. Changes in their populations account for enhanced low-pressure compressibility and configurational reorganization near the diffraction anomaly, before full proton centering.

Recent meta-GGA studies of H-bearing oxides suggest improved predictions of H-bond geometry, vibrational stability, and elasticity compared with standard GGA \cite{luoProbingStateHydrogen2024, wangInitioStudyStability2024}. Subsequent studies used the strongly constrained and appropriately normed (SCAN) meta-GGA functional \cite{sunStronglyConstrainedAppropriately2015} with deep-potential molecular dynamics (DPMD). At 300~K, SCAN-DPMD reproduces the change in the equation of state (EOS) and bulk compressibility at $\sim$10~GPa, together with changes in local O---H$\cdots$O geometry and elastic properties \cite{luoProbingStateHydrogen2024, luoElasticityAcousticVelocities2024}. Together, these studies relate spectral and compression anomalies to configurational populations and average geometry. Here we investigate the dynamical mechanisms of H-bond disordering by examining how proton-lattice coupling changes mode frequencies and damping, and why spectral coherence is lost before ordered harmonic modes soften.

Phonon quasiparticle (PHQ) analysis extracts mode-specific anharmonic frequency shifts and lifetimes from atomic motion in SCAN-DPMD \cite{zhangPhononQuasiparticlesAnharmonic2014, sunDynamicStabilizationCubic2014, zhangPBEGGAPredictsB8-B22023, zhangPhqFortranCode2019, wangPgmPythonPackage2023}. Mode-projected velocity autocorrelation functions (VAFs) record the oscillation and decay of each vibrational pattern \cite{zhangPhononQuasiparticlesAnharmonic2014, zhangPhqFortranCode2019, zhangPhononQuasiparticleApproach2022}. PHQ measures frequency shifts and correlation decay from finite-temperature atomic motion projected onto a chosen mode basis. It follows identifiable phonon peaks and their loss of coherence under pressure.

Near disorder-symmetrization at 300~K, protons sample inequivalent off-center and near-centered O---H$\cdots$O environments. These geometric variations alter the restoring forces of OH stretching and coupled optic modes. Sampling inequivalent proton environments at finite temperature can shift and broaden spectral peaks before an ordered structure becomes harmonically unstable. PHQ has also been applied to anharmonic thermodynamic and thermal transport properties of CaSiO$_3$ perovskite (CaPv) \cite{zhangInitioAnharmonicThermodynamic2021, zhangThermalConductivityCaSiO32021} and MgSiO$_3$ perovskite (MgPv) \cite{zhangInitioLatticeThermal2021, zhangAnharmonicThermodynamicProperties2022} under extreme pressures and temperatures.

Deep Potential (DP) \cite{zhangDeepPotentialMolecular2018, zhangEndendSymmetryPreserving2018} is a neural-network machine learning interatomic potential (MLIP), recently extended to hydrous minerals described with meta-GGA functionals \cite{luoProbingStateHydrogen2024, wangMachineLearningPotential2024b}. DP captures fast OH vibrations and loss of coherence near disordering through long MD simulations, at a lower cost than direct SCAN-DFT MD.

Here, we validate SCAN-DP against harmonic SCAN-DFT phonons, then compare quasiparticle frequencies and linewidths at 300~K with Raman and IR measurements. We examine the dependence of OH-stretching frequencies on H-bond geometry. For the ambient-pressure OH-stretching multiplet, we include directional longitudinal-optical--transverse-optical (LO-TO) splitting near $\Gamma$. Near disordering, individual symmetry assignments become uncertain, so we compare pressure trends in peak positions, widths, and resolvability.

\section{Methods}
\label{sec:method}

\subsection{SCAN-DPMD sampling and average structures}

We construct average structures at each pressure and 300~K to separate geometric changes from anharmonic frequency shifts. Separate trajectories supply velocities for PHQ analysis.

We performed DPMD simulations with \textsc{LAMMPS} \cite{thompsonLAMMPSFlexibleSimulation2022} using the DeepPot-SE model \cite{zhangEndendSymmetryPreserving2018} implemented in \textsc{DeePMD-kit} \cite{wangDeePMDkitDeepLearning2018, zengDeePMDkitV2Software2023}.
Our previously reported SCAN-based DeePMD potential (SCAN-DP) \cite{luoProbingStateHydrogen2024} was trained to the SCAN functional \citep{furnessAccurateNumericallyEfficient2020}.
Relative to SCAN-DFT, SCAN-DP has root-mean-square errors of $\sim$2~meV/atom in energy and 0.12~eV/\AA\ in force.

We obtained the 300~K average structures from 8,192-atom ($8 \times 8 \times 16$ supercell) DPMD simulations at constant pressure and temperature ($NPT$) with a Nos\'e-Hoover thermostat/barostat \cite{hooverKineticMomentsMethod1996}.
Each trajectory spanned 0.4~ns with a timestep of 0.2~fs.
Time averaging of primitive-cell shape and unwrapped coordinates removed thermal and barostat fluctuations without periodic-boundary artifacts.
Averaging each atom over its 1,024 translationally equivalent images reduced the supercell to an 8-atom primitive cell.

We optimized each averaged primitive cell under space-group constraints with the Atomic Simulation Environment (\textsc{ASE}) \citep{larsenAtomicSimulationEnvironment2017}.
We constrained the ordered hydrogen off-center (HOC) structures to $\mathrm{P}2_1\mathrm{nm}$ and symmetrized hydrogen-centered (HC) structures to $\mathrm{Pnnm}$ symmetry, respectively.
The $\mathrm{P}2_1\mathrm{nm}$ constraint selects HOC-1 (HOC-11 in the repeated supercell) in our earlier notation \cite{luoInitioInvestigationHbond2022}.
This arrangement is established experimentally \cite{komatsuRedeterminationHighpressureModification2006, sano-furukawaDirectObservationSymmetrization2018} and supported by \textit{ab initio} calculations \cite{luoInitioInvestigationHbond2022}.
We optimized the internal coordinates with SCAN-DP and the Broyden--Fletcher--Goldfarb--Shanno (BFGS) algorithm \citep{fletcherNewApproachVariable1970} until all residual forces were below $10^{-3}$~eV/\AA.
Lattice parameters were fixed at their $NPT$ averages; only symmetry-allowed internal coordinates relaxed.

We checked whether the symmetry-constrained optimization preserves the H-bond geometry sampled by DPMD.
At pressures with well-defined phonon quasiparticle peaks, jumps among comparable HOC motifs are infrequent in simulated proton motion at 300~K.
Optimization changes O---H and H$\cdots$O distances (Fig.~S1 of the Supplemental Material) by less than $1.2\times10^{-4}$~\AA\ for $\mathrm{P}2_1\mathrm{nm}$ and below $1.3\times10^{-2}$~\AA\ for $\mathrm{Pnnm}$.
Constrained optimization thus preserves the sampled H-bond geometry for PHQ projection.
In $\mathrm{Pnnm}$, where symmetry fixes $d(\mathrm{O}-\mathrm{H})=d(\mathrm{H}\cdots\mathrm{O})=d(\mathrm{O}\cdots\mathrm{O})/2$, the residual shifts in O---H and H$\cdots$O distances reflect variations in the O$\cdots$O separation.
Harmonic eigenvectors of the optimized average structures support branch tracking from both the off-center and centered sides of disorder-symmetrization. At each pressure, PHQ uses the corresponding optimized reference structure without averaging spectra over multiple H-bond arrangements. The DPMD trajectories are classical and do not include nuclear quantum effects.

We sampled $\mathrm{P}2_1\mathrm{nm}$ at 300~K from 0 to 18~GPa in 1~GPa increments.
For $\mathrm{Pnnm}$, we sampled from 0 to 40~GPa in 5~GPa increments, and from 60 to 140~GPa in 20~GPa increments.

We computed harmonic frequencies $\omega_{\mathbf{q}s}$ and eigenvectors for each optimized 300~K average structure. The harmonic calculations use the pressure-dependent geometry sampled by DPMD, including the axial ratios $a:b:c$ and internal coordinates that control octahedral distortions, O$\cdots$O separations, and proton off-centering. The 300~K average structures differ in cell shape from ordered 0~K structures even at the same volume (Fig.~S2 of the Supplemental Material). Force constants and eigenvectors are harmonic at each fixed structure; PHQ extracts additional anharmonic frequency shifts and linewidths from the trajectories.

Residual local distortions in directly sliced or insufficiently averaged structures produced artificial softening and imaginary harmonic branches.
Averaging and internal-coordinate optimization avoid these artifacts.

\subsection{Harmonic phonons and mode labeling}

Harmonic phonons provide eigenvectors for PHQ projection and symmetry labels for Raman and IR comparisons. We computed SCAN-DP phonons by finite displacements \cite{alfePHONProgramCalculate2009} using \textsc{PhonoPy} \cite{togoFirstPrinciplesPhonon2015, togoImplementationStrategiesPhonopy2023} and \textsc{PhonoLAMMPS} \cite{carrerasPhonoLAMMPS2023}.

To validate this harmonic description, we compared SCAN-DP with SCAN-DFT phonons computed using \textsc{VASP} \cite{kresseEfficientIterativeSchemes1996} and \textsc{PhonoPy} \cite{togoFirstPrinciplesPhonon2015, togoImplementationStrategiesPhonopy2023}. SCAN-DFT calculations used $2 \times 2 \times 4$ supercells, a $\Gamma$-centered $2 \times 2 \times 2$ $k$-point mesh, a plane-wave energy cutoff of 520~eV, the SCAN meta-GGA functional \cite{sunStronglyConstrainedAppropriately2015}, and the projector augmented-wave (PAW) method.

For Raman and IR assignments, zone-center optic modes decompose as \cite{tsuchiyaVibrationalPropertiesDAlOOH2008}
\[
\Gamma = 7 \, \mathrm{A}_1 + 4 \, \mathrm{A}_2 + 3 \, \mathrm{B}_1 + 7 \, \mathrm{B}_2
\]
for the $\mathrm{P}2_1\mathrm{nm}$ space group, and
\[
\Gamma = 2 \, \mathrm{A}_g + 3 \, \mathrm{A}_u + 2 \, \mathrm{B}_{1g} + 2 \, \mathrm{B}_{1u} + \mathrm{B}_{2g} + 5 \, \mathrm{B}_{2u} + \mathrm{B}_{3g} + 5 \, \mathrm{B}_{3u}
\]
for the $\mathrm{Pnnm}$ space group.
For electric-dipole IR activity at $\Gamma$, the allowed representations are $\mathrm{A}_1$, $\mathrm{B}_1$, and $\mathrm{B}_2$ in $\mathrm{P}2_1\mathrm{nm}$, and $\mathrm{B}_{1u}$, $\mathrm{B}_{2u}$, and $\mathrm{B}_{3u}$ in $\mathrm{Pnnm}$. The $\mathrm{A}_2$ modes are Raman active but IR inactive in $\mathrm{P}2_1\mathrm{nm}$; $\mathrm{A}_u$ modes are inactive in both first-order Raman and electric-dipole IR spectroscopy in $\mathrm{Pnnm}$.

\subsection{Non-analytical corrections near \texorpdfstring{$\Gamma$}{Gamma}}

Non-analytical corrections (NAC) are required to interpret polar near-$\Gamma$ branches because LO-TO splitting is anisotropic as $\mathbf{q} \to \mathbf{0}$ \cite{gonzeDynamicalMatricesBorn1997, baroniPhononsRelatedCrystal2001}. We include NAC when comparing directional near-zone-center frequencies with Raman and IR measurements.
For polar materials, NAC enters the phonon dynamical matrix as \cite{giannozziInitioCalculationPhonon1991, gonzeDynamicalMatricesBorn1997, baroniPhononsRelatedCrystal2001}
\begin{equation}
    \tilde{D}_{\kappa\alpha, \kappa'\beta}^{\mathrm{NA}}(\mathbf{q}) =
    \frac{4\pi}{\Omega_0 \sqrt{M_\kappa M_{\kappa'}}}
    \frac{\left(\sum_\gamma q_\gamma Z_{\kappa,\gamma\alpha}^*\right)\left(\sum_{\gamma'} q_{\gamma'} Z_{\kappa',\gamma'\beta}^*\right)}{\sum_{\alpha\beta} q_\alpha \, \epsilon_{\alpha\beta}^{\infty} \, q_\beta} ,
\label{eq:nac}
\end{equation}
where $\Omega_0$ is the primitive-cell volume, $M_\kappa$ is the mass of atom $\kappa$, $Z^*_{\kappa,\gamma\alpha}$ is the Born effective charge tensor, and $\epsilon^\infty_{\alpha\beta}$ is the electronic dielectric tensor. Born charges projected along $\mathbf{q}$ set the coupling to the long-range electric field, screened by the dielectric tensor in the denominator.

At each pressure, we computed the Born effective charge and electronic dielectric tensors entering Eq.~\eqref{eq:nac} with \textsc{VASP} \cite{kresseEfficientIterativeSchemes1996} on the 8-atom primitive cell with a $\Gamma$-centered $4 \times 4 \times 4$ $k$-point mesh. These tensors determine NAC for both harmonic and quasiparticle phonons near $\Gamma$.
Near-zone-center frequencies thus depend on propagation direction in Raman and IR comparisons.

\subsection{Phonon quasiparticle extraction}
\label{sec:phq}

Phonon quasiparticle analysis determines mode-specific frequencies and correlation decay from finite-temperature atomic motion. At each pressure, we project velocities from separate constant-volume and temperature ($NVT$) trajectories onto harmonic eigenvectors of the optimized 300~K structures. Symmetry constraints apply to the average structures; production dynamics samples the full SCAN-DP potential without a harmonic or low-order force-constant truncation. Harmonic-mode projection decomposes the anharmonic atomic motion into these displacement patterns \cite{zhangPhononQuasiparticlesAnharmonic2014, zhangPhqFortranCode2019}. Anharmonic interactions can mix the patterns even with pressure-specific eigenvectors. Individual quasiparticle contributions are identified from the projected peak shapes.

To resolve fast OH oscillations and rapidly decaying correlations, we performed $NVT$ DPMD simulations with a Nos\'e-Hoover thermostat \cite{hooverKineticMomentsMethod1996}, using $4 \times 4 \times 4$ supercells (512 atoms). All runs used a timestep of 0.2~fs and at least 10,000 equilibration steps before production. Near the low-pressure anomaly, production lasted $2 \times 10^6$ steps (0.4~ns), with velocities saved every 10 steps. For the coarser high-pressure grid, production lasted $4 \times 10^5$ steps (80~ps), with velocities saved every step. We analyzed all trajectories with \textsc{DynaPhoPy} \cite{zhangPhqFortranCode2019, carrerasDynaPhoPyCodeExtracting2017}.

For branch $s$ at a wave vector $\mathbf{q}$ commensurate with the MD supercell, the mode-projected velocity is \cite{carrerasDynaPhoPyCodeExtracting2017}
\begin{equation}
    V_{\mathbf{q}s}(t) = \frac{1}{\sqrt{N_c}}
    \sum_{l\kappa} \sqrt{M_\kappa}\,
    \mathbf{e}_{\kappa s}^{*}(\mathbf{q})\cdot\mathbf{v}_{l\kappa}(t)
    e^{-i\mathbf{q}\cdot\mathbf{r}^{0}_{l\kappa}} .
    \label{eq:projected-velocity}
\end{equation}
Here, $N_c$ is the number of primitive cells in the MD supercell, $l$ labels those cells, and $\kappa$ labels atoms within a primitive cell. $\mathbf{v}_{l\kappa}(t)$ is the atomic velocity, $M_\kappa$ the atomic mass, $\mathbf{r}^{0}_{l\kappa}$ the atomic position in the optimized cell repeated over the supercell, and $\mathbf{e}_{\kappa s}(\mathbf{q})$ the harmonic eigenvector with $\sum_\kappa |\mathbf{e}_{\kappa s}(\mathbf{q})|^2=1$. Spatial phases select the wave vector, eigenvector projection resolves the displacement pattern, and $N_c^{-1/2}$ normalizes the sum over equivalent cells.

The mode-projected VAF is \cite{wangTightbindingMoleculardynamicsStudy1990, carStructuralDymanicalElectronic1988, zhangPhononQuasiparticlesAnharmonic2014, zhangPhqFortranCode2019}
\begin{equation}
    \mathrm{VAF}
    = \left\langle V_{\mathbf{q}s}^{*}(0) V_{\mathbf{q}s}(t)\right\rangle
    = \lim _{\mathcal{T} \rightarrow \infty} \frac{1}{\mathcal{T}} \int_0^{\mathcal{T}} V_{\mathbf{q}s}^*(t') \, V_{\mathbf{q}s}(t' + t) \, dt' ,
    \label{eq:vaf}
\end{equation}
where $\langle \cdot \rangle$ denotes a trajectory average over the interval $\mathcal{T}$. The VAF measures the persistence of motion along the selected displacement pattern. For a well-defined quasiparticle and $t \geq 0$, it has the damped-cosine form \cite{sunPhononQuasiparticlesAnharmonic2010, zhangPhononQuasiparticlesAnharmonic2014, zhangPhqFortranCode2019}
\begin{equation}
    \left\langle V_{\mathbf{q}s}^{*}(0)V_{\mathbf{q}s}(t) \right\rangle
    \simeq A_{\mathbf{q}s}\cos(\tilde\omega_{\mathbf{q}s}t)
    e^{-t/(2\tau_{\mathbf{q}s})},
    \label{eq:vaf-decay}
\end{equation}
where $A_{\mathbf{q}s}$ is the correlation amplitude and $\tau_{\mathbf{q}s}$ is the quasiparticle relaxation lifetime. The correlation envelope decays on the timescale $2\tau_{\mathbf{q}s}$. Fourier transformation defines the mode-projected power spectrum \cite{zhangPhononQuasiparticlesAnharmonic2014, zhangPhqFortranCode2019},
\begin{equation}
G_{\mathbf{q}s}(\omega) = \int_{-\infty}^{\infty}\left\langle V_{\mathbf{q}s}^{*}(0) V_{\mathbf{q}s}(t)\right\rangle e^{i \omega t} d t .
    \label{eq:power-spectrum}
\end{equation}
The damped oscillation produces a Lorentzian peak in $G_{\mathbf{q}s}(\omega)$. Lorentzian fitting with \textsc{DynaPhoPy} determines the quasiparticle frequencies $\tilde\omega_{\mathbf{q}s}$ and linewidths $\Gamma_{\mathbf{q}s}$. We report linewidths as full widths at half maximum (FWHM) in cm$^{-1}$.

We tracked each branch through the disordering interval from the low- or high-pressure side while $G_{\mathbf{q}s}(\omega)$ exhibited one dominant, approximately Lorentzian peak. Branch tracking ended when comparable secondary peaks or spectral overlap prevented a unique assignment. Linewidths approaching the peak frequency also limit the quasiparticle description. In this regime, broadening and reduced peak resolvability characterize the Raman and IR anomalies, while individual frequencies, lifetimes, and symmetry assignments remain ambiguous.

\section{Results and Discussion}
\label{sec:results}

\subsection{Validation of harmonic phonons}

\begin{figure}[!htbp]
    \centering
    \includegraphics[width=\columnwidth]{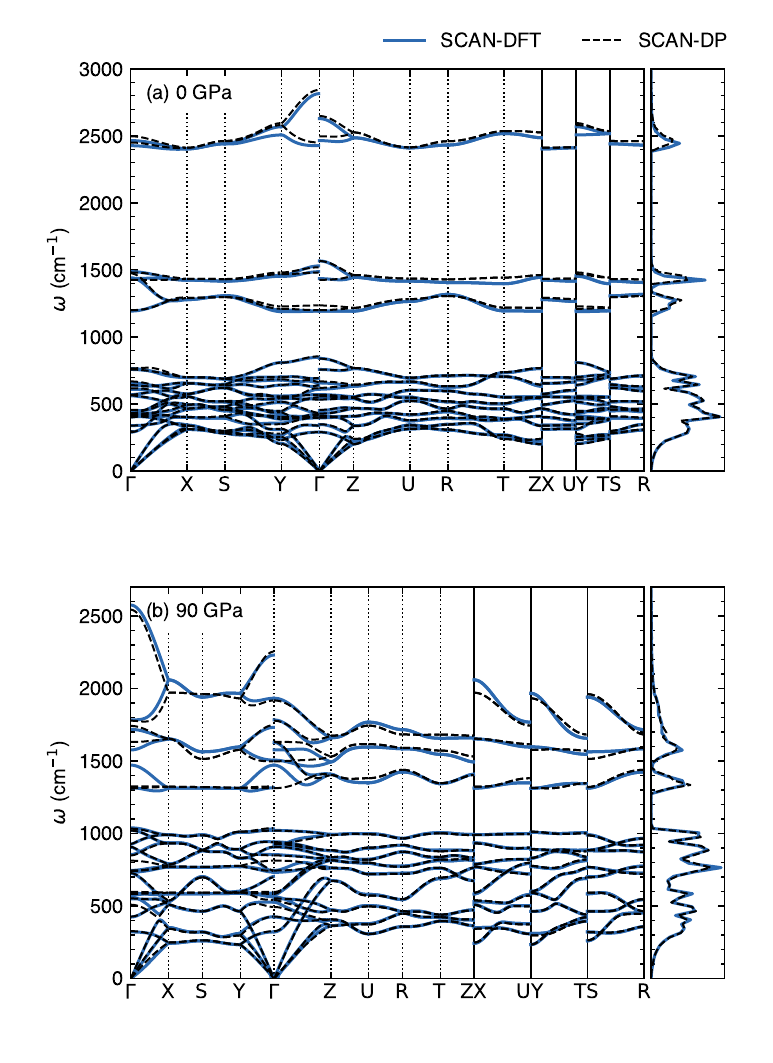}
    \caption{Harmonic phonon dispersion and VDoS for $\delta$-AlOOH from SCAN-DFT and SCAN-DP at (a)~0~GPa and (b)~90~GPa.}
    \label{fig:dft}
\end{figure}

Finite-temperature analysis requires an accurate harmonic description of framework and OH vibrations. Our previous study established agreement between SCAN-DP and SCAN-DFT energetics, forces, and structures of $\delta$-AlOOH \cite{luoProbingStateHydrogen2024}.

Figure~\ref{fig:dft} compares harmonic dispersions and vibrational densities of states (VDoS) at 0~GPa ($\mathrm{P}2_1\mathrm{nm}$) and 90~GPa ($\mathrm{Pnnm}$), using static optimized structures.
Both SCAN-DP dispersions are stable and closely match SCAN-DFT, except for minor high-frequency differences. SCAN-DP and SCAN-DFT VDoS curves nearly coincide. This agreement supports using SCAN-DP harmonic phonons to assess finite-temperature frequency shifts.

Under static conditions, SCAN-DP reproduces the qualitative pressure dependence previously reported for PBE-DFT/density-functional-perturbation-theory (DFPT) calculations. Static SCAN-DP predicts stable zone-center modes through the experimental 5--10~GPa anomaly range, with softening only toward symmetrization at $\sim$30~GPa (Fig.~S4 of the Supplemental Material), above the observed Raman and IR anomaly pressures \cite{tsuchiyaVibrationalPropertiesDAlOOH2008}. Ordered harmonic phonons thus miss the early anomaly even when the potential reproduces the DFT phonons.

Unless a reciprocal-space direction is specified, zone-center frequencies below refer to unsplit branches; directional LO-TO splitting is treated separately.

\subsection{Quasiparticle renormalization across the Brillouin zone}

\begin{figure}[htbp]
    \centering
    \includegraphics[width=\columnwidth]{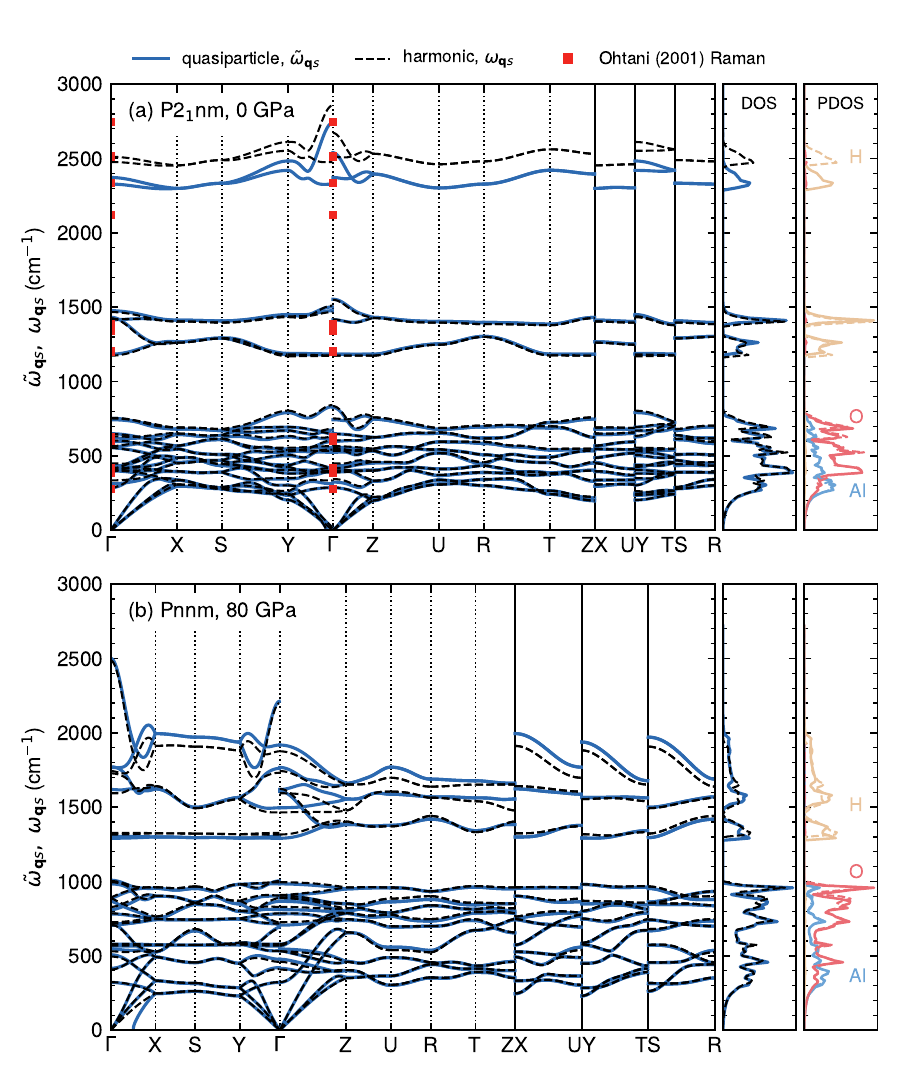}
    \caption{Phonon dispersions and total and species-projected VDoS at 300~K for $\delta$-AlOOH at (a)~0~GPa ($\mathrm{P}2_1\mathrm{nm}$) and (b)~80~GPa ($\mathrm{Pnnm}$). Solid blue and black dashed curves show quasiparticle frequencies $\tilde\omega_{\mathbf{q}s}$ and harmonic frequencies $\omega_{\mathbf{q}s}$ of the corresponding 300~K average structures, respectively. VDoS projections resolve Al, O, and H contributions. Red squares mark ambient-pressure Raman peaks at 300~K \cite{ohtaniStabilityFieldNew2001}.}
    \label{fig:2}
\end{figure}

At 300~K, framework and OH vibrations respond differently to compression and anharmonicity (Fig.~\ref{fig:2}). Al motion dominates the lowest-frequency modes, with O participation in octahedral motion. At intermediate frequencies, framework vibrations involve predominantly O motion, while H motion dominates OH bending and stretching above $\sim$1,200~cm$^{-1}$. Compression to 80~GPa raises the O-dominated frequencies and lowers the H-dominated frequencies, narrowing the gap between framework and OH-related vibrations, consistent with stronger hydrogen bonding.

At 0~GPa, anharmonic frequency shifts are largest for H-dominated vibrations and small for many framework modes. H-related branches also show the clearest renormalization at 80~GPa, although the shifts vary in sign and magnitude among modes.

\subsection{Ambient-pressure OH-stretching Raman comparison}

Four broad ambient-pressure OH-stretching Raman peaks span a wider frequency range than the two unsplit stretching modes at 300~K \cite{ohtaniStabilityFieldNew2001}. Assessing the multiplet requires directional near-$\Gamma$ frequencies.

Quasiparticle renormalization lowers OH-stretching frequencies across the Brillouin zone by $\sim$140--150~cm$^{-1}$ at ambient pressure. The lower-frequency stretching mode approaches the second-lowest Raman peak, while both modes lie below the two highest-frequency components. Renormalization alone gives a narrower frequency spread than observed.

Directional LO-TO splitting extends the higher-frequency stretching mode across the two highest Raman components. A $\sim$370~cm$^{-1}$ upshift along $\mathrm{Y} \rightarrow \Gamma$ approaches the highest peak. Along $\Gamma \rightarrow \mathrm{X}$, LO-TO splitting barely shifts the lower-frequency mode.

Together, renormalization and directional splitting reproduce the frequencies of the three higher Raman components. The unresolved component at $\sim$2,121~cm$^{-1}$ is consistent with configurational splitting beyond the present PHQ description \cite{luoInitioInvestigationHbond2022}.

Polycrystalline Raman samples include different propagation and polarization directions through their grain orientations \cite{ohtaniStabilityFieldNew2001, mashinoSoundVelocitiesDAlOOH2016}. Figure~\ref{fig:2} illustrates the resulting frequency spread without a full orientational average; mode-resolved frequencies and linewidths are given in Table~S1 of the Supplemental Material.

\subsection{Evolution of vibrational anomalies under pressure}

\begin{figure*}
    \centering
    \includegraphics[width=.9\textwidth]{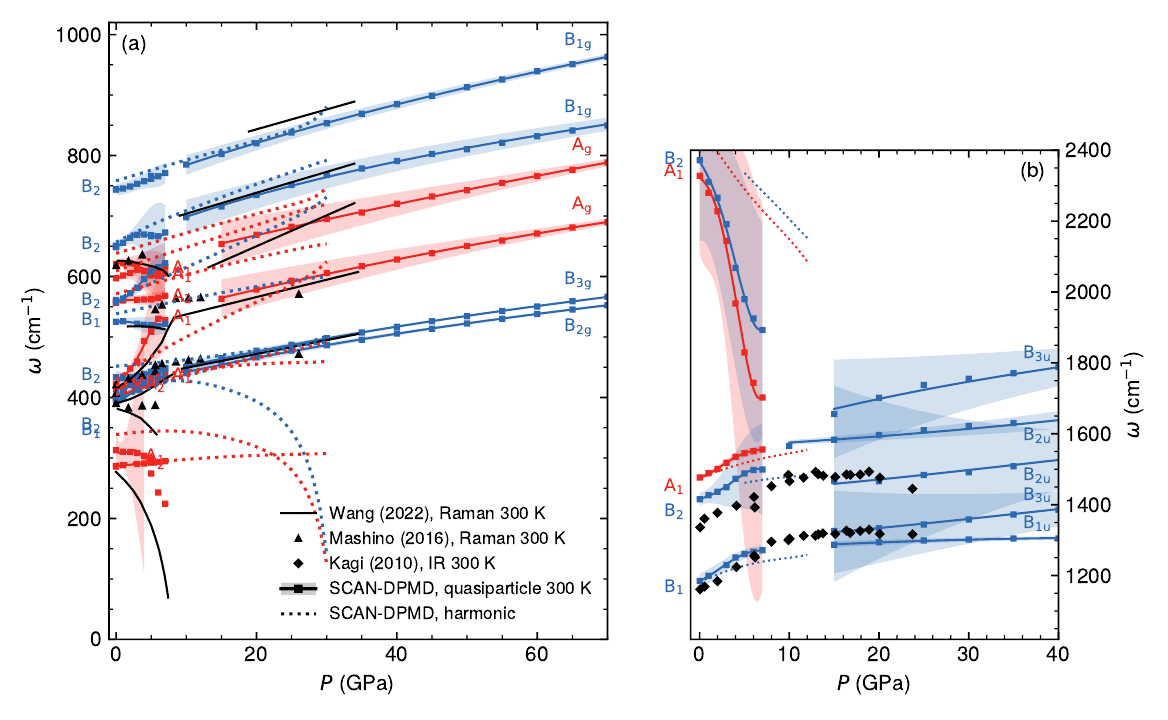}
    \caption{Harmonic and quasiparticle frequencies, $\omega_{\mathbf{q}s}$ and $\tilde\omega_{\mathbf{q}s}$, for selected zone-center modes at 300~K. (a) Raman-active modes below 1,000~cm$^{-1}$. (b) IR-active modes above 1,100~cm$^{-1}$. Dotted curves show the harmonic frequencies, which are finite throughout the experimental 5--10~GPa anomaly interval and soften only toward much higher pressure, consistent with earlier static \textit{ab initio} pressure trends \cite{tsuchiyaVibrationalPropertiesDAlOOH2008}. Solid curves and symbols show quasiparticle frequencies; shading spans $\tilde\omega_{\mathbf{q}s} \pm \Gamma_{\mathbf{q}s}/2$, with full width $\Gamma_{\mathbf{q}s}$. In (b), red and blue denote $\mathrm{A}$- and $\mathrm{B}$-type branches, respectively. Raman and IR measurements \cite{wangRamanScatteringCr32022, mashinoSoundVelocitiesDAlOOH2016, kagiInfraredAbsorptionSpectra2010}.}
    \label{fig:raman}
\end{figure*}

At 300~K, peak shifts, linewidth growth, and branch reorganization mark the early vibrational stage of disordering. We compare their onset pressures with those of the bulk compression and average-symmetry changes. Figure~\ref{fig:raman} compares measured peaks with quasiparticle frequencies $\tilde\omega_{\mathbf{q}s}$ and linewidths $\Gamma_{\mathbf{q}s}$ \cite{wangRamanScatteringCr32022, kagiInfraredAbsorptionSpectra2010, mashinoSoundVelocitiesDAlOOH2016}. Plotted frequencies omit directional LO-TO splitting.

Quasiharmonic frequencies are positive throughout the experimental 5--10~GPa anomaly interval. Several low-frequency Raman-active branches develop broad minima only at $\sim$30~GPa, consistent with earlier static \textit{ab initio} results \cite{tsuchiyaVibrationalPropertiesDAlOOH2008}. The low-frequency $\mathrm{A}_1$ mode softens gradually, while the branches above 1,000~cm$^{-1}$ undergo avoided crossings.

Raman and IR measurements under controlled pressure provide more reliable constraints on the anomaly onset \cite{wangRamanScatteringCr32022, kagiInfraredAbsorptionSpectra2010}, whereas the polycrystalline Raman study without comparable stress control provides a qualitative comparison for the appearance, disappearance, and broadening of peaks \cite{mashinoSoundVelocitiesDAlOOH2016}. Near the disordering regime, LO-TO splitting, linewidth growth, and branch reordering limit one-to-one mode assignment, so the most direct comparison is through pressure trends, peak positions, linewidths, and peak resolution.

Low-frequency Raman anomalies involve an $\mathrm{A}_1$ mode from 0 to 3~GPa [Fig.~\ref{fig:raman}(a)]. Assignments above 4~GPa become uncertain as low-frequency spectral peaks broaden and overlap, with the linewidth reaching the scale of the oscillation frequency at $\sim$5~GPa. Loss of vibrational coherence is examined in Sec.~\ref{sec:interpretation}.

Polycrystalline Raman measurements likewise show strong low-frequency peaks broadening or disappearing above $\sim$5.6~GPa, with new higher-frequency peaks emerging \cite{mashinoSoundVelocitiesDAlOOH2016}.

Figure~\ref{fig:raman}(b) compares two measured IR peaks below $\sim$1,500~cm$^{-1}$. Quasiharmonic OH-stretching and bending branches approach one another through avoided crossings \cite{tsuchiyaVibrationalPropertiesDAlOOH2008}. Quasiparticle branch reorganization occurs at lower pressure.

Under compression, the lower-frequency IR peak hardens up to disordering and then plateaus over 12--24~GPa \cite{kagiInfraredAbsorptionSpectra2010}. The IR-active $\mathrm{B}_1$ branch follows the initial hardening, while $\mathrm{B}_{3u}$ best matches the high-pressure plateau. Mode assignments thus change across the crossover.

For the higher-frequency IR peak, hardening and flattening correspond to $\mathrm{B}_2$ at low pressure and $\mathrm{B}_{2u}$ at high pressure \cite{kagiInfraredAbsorptionSpectra2010}.

Finite-temperature renormalization improves the $\mathrm{B}_2$ bending pressure trend. The low-pressure hardening rate increases from $\sim$6~cm$^{-1}$/GPa in the harmonic calculation on the 300~K average structures to $\sim$13~cm$^{-1}$/GPa for PHQ over 0--7~GPa, close to the measured rate of $\sim$12~cm$^{-1}$/GPa below 8~GPa \cite{kagiInfraredAbsorptionSpectra2010}. Quasiparticle $\mathrm{B}_2$ frequencies also flatten at higher pressure, capturing the measured trend despite residual absolute-frequency offsets.

Figure~\ref{fig:raman}(b) omits directional LO-TO splitting, which introduces uncertainty in the bending-frequency comparison. At 0~GPa and 300~K, directional LO-TO splitting is $\sim$86~cm$^{-1}$ for $\mathrm{B}_2$ bending along $b^*$ and $\sim$77~cm$^{-1}$ for $\mathrm{A}_1$ bending along $c^*$, relative to their unsplit $\Gamma$ frequencies. Directional splitting exceeds the anharmonic shifts of these branches and may contribute substantially to the absolute-frequency offsets. This comparison is limited to principal propagation directions.

At high pressure, peak separation restores identifiable modes. One $\mathrm{B}_{2u}$ branch emerges above $\sim$10~GPa, while several overlapping low-frequency peaks separate only above $\sim$15~GPa.

At 300~K, the vibrational changes begin before the bulk compression and average-symmetry anomalies. The low-frequency $\mathrm{A}_1$ vibration loses coherence at $\sim$4--5~GPa, and Raman peaks broaden or disappear above $\sim$5.6~GPa \cite{mashinoSoundVelocitiesDAlOOH2016}. Larger branch reorganization and frequency shifts develop across 5--10~GPa. Neutron diffraction identifies equivalent off-center occupations and $\mathrm{Pnnm}$ average symmetry at $\sim$9~GPa \cite{sano-furukawaDirectObservationSymmetrization2018}, while SCAN-DPMD gives the bulk compression anomaly at $\sim$10~GPa \cite{luoProbingStateHydrogen2024, luoElasticityAcousticVelocities2024}. The spectral onset marks the early vibrational stage of H-bond disordering.

At higher pressure, many branches sharpen and their frequencies approach the harmonic frequencies. Several low-frequency peaks become distinguishable again above $\sim$15~GPa. Vibrational coherence recovers as the H-bond geometry approaches proton centering at $\sim$16--18~GPa. Raman and luminescence measurements likewise show subtler changes at centering than in the earlier disordering interval \cite{wangRamanScatteringCr32022}.

\subsection{Physical interpretation}
\label{sec:interpretation}

Raman spectra probe coupled proton-lattice coherence, bulk compressibility describes how mean volume changes with pressure, and diffraction measures proton-site occupations and positions. Coherence, compressibility, and site occupations change at different pressures during H-bond reorganization; the earliest vibrational change is loss of coherence in the mode examined below.

\begin{figure*}[htbp]
    \centering
    \includegraphics[width=.83\textwidth]{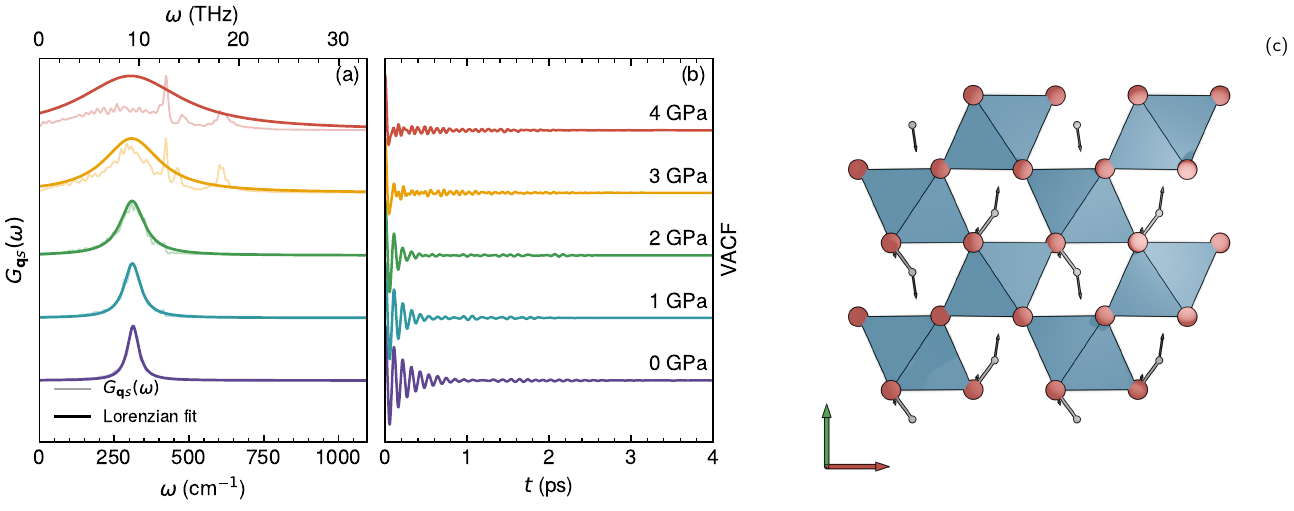}

    \caption{Mode-projected spectra and velocity correlations at $\Gamma$ and 300~K for the low-frequency branch evolving from the 0~GPa $\mathrm{A}_1$ mode. (a) Projected power spectra, $G_{\mathbf{q}s}(\omega)$, and single-Lorentzian fits from 0 to 4~GPa. (b) Mode-projected velocity autocorrelation functions for the same pressures. (c) Top view of the 0~GPa $\mathrm{P}2_1\mathrm{nm}$ mode pattern. Translucent blue polyhedra denote $\mathrm{AlO}_6$ octahedra, red spheres O atoms, and small white spheres H atoms; gray arrows indicate the displacement directions. Al atoms lie at octahedral centers. Displacement amplitudes are exaggerated for visibility, preserving their ratios among atoms. Over 0--4~GPa, the linewidth grows strongly while the peak position changes little, indicating loss of coherence before substantial frequency softening.}
    \label{fig:soft-spectra}
\end{figure*}

Linewidth growth measures the loss of vibrational coherence. Following Eq.~\eqref{eq:vaf-decay} in Sec.~\ref{sec:phq}, the relaxation lifetime is related to the FWHM $\Gamma_{\mathbf{q}s}$ in cm$^{-1}$ by
\begin{equation}
    \tau_{\mathbf{q}s} \simeq \left(2\pi c\,\Gamma_{\mathbf{q}s}\right)^{-1},
    \label{eq:linewidth-lifetime}
\end{equation}
where $c$ is the speed of light. Broader peaks, therefore, correspond to shorter relaxation lifetimes.

Between 0 and 4~GPa at 300~K, the low-frequency $\mathrm{A}_1$ peak broadens from $\sim$50 to 400~cm$^{-1}$ while its central frequency changes little (Fig.~\ref{fig:soft-spectra}). The single Lorentzian peak at ambient pressure develops secondary structure under compression. By 4~GPa, the width is comparable to the central frequency, and the spectral weight extends toward zero frequency. The fitted width then characterizes spectral broadening; its conversion to a quasiparticle lifetime is ambiguous. This loss of coherence precedes the harmonic instability of the ordered structure.

Previous Raman measurements assigned the soft $\mathrm{A}_1$ mode to an $\mathrm{AlO}_6$-octahedral vibration coupled to OH stretching \cite{wangRamanScatteringCr32022}. At 0~GPa, the calculated eigenvector is polarized almost entirely along $b$: the two Al atoms move out of phase, neighboring octahedra along the H-bond tunnel counter-rotate, and H atoms move with the local framework [Fig.~\ref{fig:soft-spectra}(c)] \cite{vanpeteghemNeutronDiffractionStudy2007}. Framework rotation, Al translation, and proton displacement form a single coupled normal coordinate whose relative contributions change little over 0--4~GPa. Proton off-centering changes the local restoring-force balance within this coupled motion.

Octahedral counter-rotation resembles the stishovite $\mathrm{B}_{1g}$ branch and its post-stishovite $\mathrm{A}_g$ continuation, which couple octahedral motion to ferroelastic strain \cite{togoFirstprinciplesCalculationsFerroelastic2008, zhangElasticityPseudoproperFerroelastic2021}. Across the $\mathrm{P}4_2/\mathrm{mnm}\rightarrow\mathrm{Pnnm}$ distortion, the SiO$_2$ eigenvector retains an almost pure in-plane O-framework rotation with negligible Si displacement. Al and H motion in $\delta$-AlOOH makes its $\mathrm{A}_1$ mode a coupled proton-lattice antiferrodistortive vibration. In SiO$_2$, the analogous branch is the static harmonic soft mode of the ordered ferroelastic transition; the 300~K anomaly in $\delta$-AlOOH begins with linewidth growth and loss of coherence before ordered harmonic softening.

Our earlier multiconfiguration study found H-bond arrangements ordered in the $ab$ plane but differing in stacking along $c$ \cite{luoInitioInvestigationHbond2022}. The mixed HOC-12 arrangement and the half-aligned HOC-11* arrangement have different OH-stretching spectra, free energies, and volumes. Changes in their populations accounted for enhanced low-pressure compressibility and reorganization near 8--9~GPa on the experimental pressure scale. PHQ complements this configurational picture by resolving the damping of coupled framework-proton vibrations. The eigenvector and correlation decay in Fig.~\ref{fig:soft-spectra} connect the Raman anomaly to loss of coherence in this collective coordinate.

Local variations in proton off-centering change the restoring forces within the coupled framework-proton coordinate. Anharmonic interactions can shorten its correlation time before the mean structure changes substantially. The bulk compression anomaly develops as the average geometry and the H-bond network's resistance to compression change. Linewidth growth and Raman-peak disappearance can therefore occur at different pressures from an EOS slope change or equal proton-site occupations. The different pressures are consistent with the multistage configurational picture \cite{luoInitioInvestigationHbond2022} and the diffraction sequence \cite{sano-furukawaDirectObservationSymmetrization2018}.

In compressed ice, Komatsu \textit{et al.} distinguished spatially disordered, hopping, tunneling, and centered proton states within the evolution from ice VII to ice X \cite{komatsuHydrogenBondSymmetrisation2024}. Yamane, Komatsu, and Kagi identified a crossover from molecular rotation to proton translation in ice VII through their different relaxation times \cite{yamaneDirectEvidenceProtonDynamics2021}. Dynamical observables can thus distinguish changes within a structurally disordered regime. In $\delta$-AlOOH, our projected spectra follow the loss and recovery of coherent proton-lattice vibrations during disorder-symmetrization.

\subsection{OH-stretching frequencies and H-bond geometry}

\begin{figure*}[htbp]
    \centering
    \includegraphics[width=.80\textwidth]{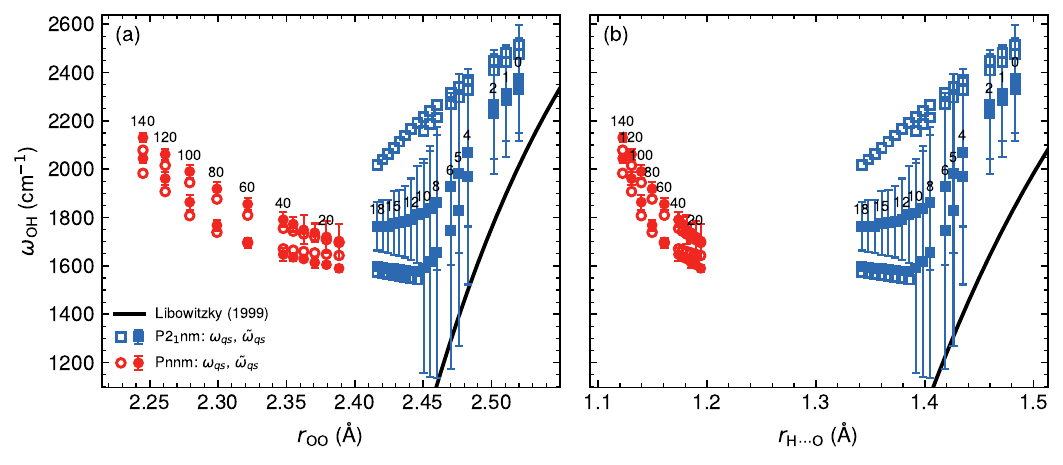}
    \caption{Zone-center OH-stretching frequencies versus H-bond geometry at 300~K. (a) Frequency versus average O$\cdots$O distance, $r_{\mathrm{OO}}$. (b) Frequency versus average H$\cdots$O distance, $r_{\mathrm{H\cdots O}}$. Squares and circles denote $\mathrm{P}2_1\mathrm{nm}$ and $\mathrm{Pnnm}$, respectively. Open and filled symbols show harmonic frequencies $\omega_{\mathbf{q}s}$ and quasiparticle frequencies $\tilde\omega_{\mathbf{q}s}$, respectively, with error bars on quasiparticle frequencies spanning $\tilde\omega_{\mathbf{q}s} \pm \Gamma_{\mathbf{q}s}/2$. Pressures are labeled in GPa. Black curves show empirical $\nu$--$d(\mathrm{O}\cdots\mathrm{O})$ and $\nu$--$d(\mathrm{H}\cdots\mathrm{O})$ correlation analyses for hydrogen-bearing minerals \cite{libowitzkyCorrelationOHStretching1999}.}
    \label{fig:hbond-stretch}
\end{figure*}

OH-stretching frequencies reflect the restoring forces set by local H-bond geometry. Figure~\ref{fig:hbond-stretch} compares zone-center OH-stretching frequencies with empirical frequency--distance correlations at the mean 300~K O$\cdots$O and H$\cdots$O distances \cite{libowitzkyCorrelationOHStretching1999}. Additional projected spectra are given in Fig.~S3 of the Supplemental Material.

Quasiparticle renormalization lowers OH-stretching frequencies at low pressure toward the empirical curves, improving agreement in both frequency and pressure dependence despite systematic offsets. Across the symmetry crossover, the lower-frequency quasiparticle OH-stretching mode is nearly continuous despite the abrupt change in mean H-bond geometry.

Directional LO-TO splitting, omitted in Fig.~\ref{fig:hbond-stretch}, and inequivalent local H-bond environments can contribute to residual OH-stretching offsets \cite{tsuchiyaVibrationalPropertiesDAlOOH2008, luoInitioInvestigationHbond2022}. Errors in the electronic-structure functional may also contribute.

Earlier multiconfiguration calculations reproduced the ambient Raman multiplet after including inequivalent local H-bond arrangements. Half-aligned proton arrangements contributed lower-frequency OH-related components than fully aligned ones \cite{luoInitioInvestigationHbond2022}. Quantum vibrational free energies, including zero-point energy, also changed the relative populations of the H-bond arrangements in the multiconfiguration quasiharmonic calculation \cite{luoInitioInvestigationHbond2022}. Classical dynamics omits these quantum population changes, which could alter relative spectral weights. Quantum shifts of individual frequencies require separate assessment. Our results cannot separate errors from PHQ, incomplete configurational sampling, and omitted nuclear quantum effects.

Raman spectra from polarizability correlations could test PHQ peak positions and broadening. Total IR or dielectric spectra have been calculated from dipole or polarization correlations in water, ice, molecules, and PbTiO$_3$ \cite{zhangDeepNeuralNetwork2020, bhatiaMACE4IRFoundationModel2025, bhatiaMACE4IRmolUncertaintyawareFoundation2025, xieThermalDisorderPhonon2025}. Directional LO-TO splitting requires a separate treatment of the long-range electric field. PHQ provides the branch-resolved frequencies and linewidths needed to interpret the total spectra.

\section{Conclusions}
\label{sec:conclusion}

At 300~K, PHQ provides insight into the dynamical mechanisms of H-bond disordering in $\delta$-AlOOH through coupled proton-lattice vibrations. The low-frequency Raman anomaly begins with linewidth growth and loss of coherence at $\sim$4--5~GPa while its characteristic frequency is finite. Branch reorganization and reduced peak resolution across 5--10~GPa account for the spectroscopic anomalies that ordered harmonic phonons miss. Higher-pressure peaks sharpen, with more gradual changes in frequencies and linewidths near proton centering at $\sim$16--18~GPa.

At ambient pressure, quasiparticle renormalization and directional near-$\Gamma$ LO-TO splitting together reproduce the three higher-frequency components of the OH-stretching Raman multiplet. The $\sim$2,121~cm$^{-1}$ component requires an explanation beyond this treatment. Quasiparticle OH-stretching frequencies also approach the empirical frequency--geometry trends compiled across hydrogen-bearing minerals \cite{libowitzkyCorrelationOHStretching1999}, with residual offsets.

The different onset pressures for loss of vibrational coherence, anomalous bulk compression, and equal proton-site occupations are consistent with multistage H-bond disordering. Finite-temperature H-bond fluctuations and strong anharmonic coupling are necessary to explain the measured spectra. Future work will examine proton spatial correlations and intersite motion, and assess how configurational averaging and nuclear quantum effects contribute to the remaining OH-stretching discrepancies.

\begin{acknowledgments}
This research was initiated under the support of DOE Award DE-SC0019759. It has also been supported by the Gordon \& Betty Moore Foundation Award GBMF12801 (\url{doi.org/10.37807/GBMF12801}) \cite{gordonandbettymoorefoundationModelingEarthAtomic2024} and NSF Award EAR 25-06448. It used the \textit{Delta} system at the National Center for Supercomputing Applications (NCSA) through allocation TG-DMR180081 from the Advanced Cyberinfrastructure Coordination Ecosystem: Services \& Support (ACCESS) program, which is supported by U.S.\ National Science Foundation grants \#2138259, \#2138286, \#2138307, \#2137603, and \#2138296.
\end{acknowledgments}

\bibliography{Geophysics}

\end{document}

\title{Supplementary material for ``Lattice dynamics and the spectroscopic signatures of H-bond disorder in \texorpdfstring{$\delta$}{delta}-AlOOH"}

\author{Chenxing Luo\,\orcidlink{0000-0003-4116-6851}}
\email{cl3658@columbia.edu}
\affiliation{Department of Applied Physics and Applied Mathematics, Columbia University, New York, New York 10027, USA}
\affiliation{Department of Geosciences, Princeton University, Princeton, New Jersey 08544, USA}

\author{Sangjoon Lee\,\orcidlink{0000-0002-2367-3932}}
\altaffiliation[Current address: ]{Department of Materials Science and Engineering, Stanford University, Stanford, California 94305, USA}
\affiliation{Department of Applied Physics and Applied Mathematics, Columbia University, New York, New York 10027, USA}

\author{Hongjin Wang\,\orcidlink{0009-0008-9766-1177}}
\affiliation{Department of Applied Physics and Applied Mathematics, Columbia University, New York, New York 10027, USA}

\author{Zhen Zhang\,\orcidlink{0009-0001-0810-8054}}
\affiliation{Department of Physics and Astronomy, Iowa State University, Ames, Iowa 50011, USA}

\author{Renata Wentzcovitch\,\orcidlink{0000-0001-5663-9426}}
\email{rmw2150@columbia.edu}
\affiliation{Department of Applied Physics and Applied Mathematics, Columbia University, New York, New York 10027, USA}
\affiliation{Department of Earth and Environmental Sciences, Columbia University, New York, New York 10027, USA}
\affiliation{Lamont--Doherty Earth Observatory, Columbia University, Palisades, New York 10964, USA}
\affiliation{Data Science Institute, Columbia University, New York, New York 10027, USA}

\maketitle

\section*{This PDF file includes:}

\begin{itemize}
\item Figures~S1 to S4
\item Table S1
\end{itemize}

\newpage

\section{Supplementary figures}

\begin{figure}[htbp]
    \centering
    \includegraphics[width=.60\textwidth]{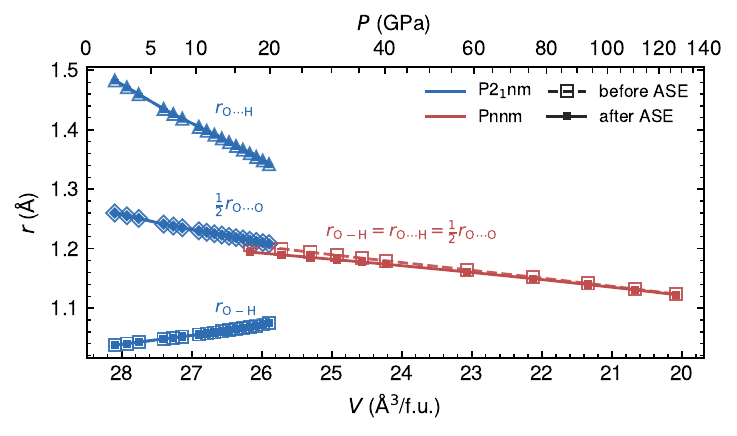}
    \caption{Hydrogen-bond geometry in the 300~K average structures before and after the final space-group-constrained internal-coordinate optimization. Distances are plotted against volume per $\delta$-AlOOH; the upper axis gives the corresponding pressure from the optimized structures. Open, larger symbols with dashed lines denote structures before optimization; filled, smaller symbols with solid lines denote the optimized structures. For $\mathrm{P}2_1\mathrm{nm}$, $d(\mathrm{O}-\mathrm{H})$, $d(\mathrm{H}\cdots\mathrm{O})$, and $d(\mathrm{O}\cdots\mathrm{O})/2$ are shown separately. For $\mathrm{Pnnm}$, one representative distance is shown because symmetry enforces $d(\mathrm{O}-\mathrm{H}) = d(\mathrm{H}\cdots\mathrm{O}) = d(\mathrm{O}\cdots\mathrm{O})/2$.}
    \label{fig:supp-average-structure-hbond}
\end{figure}

\begin{figure}[htbp]
    \centering
    \includegraphics[width=.60\textwidth]{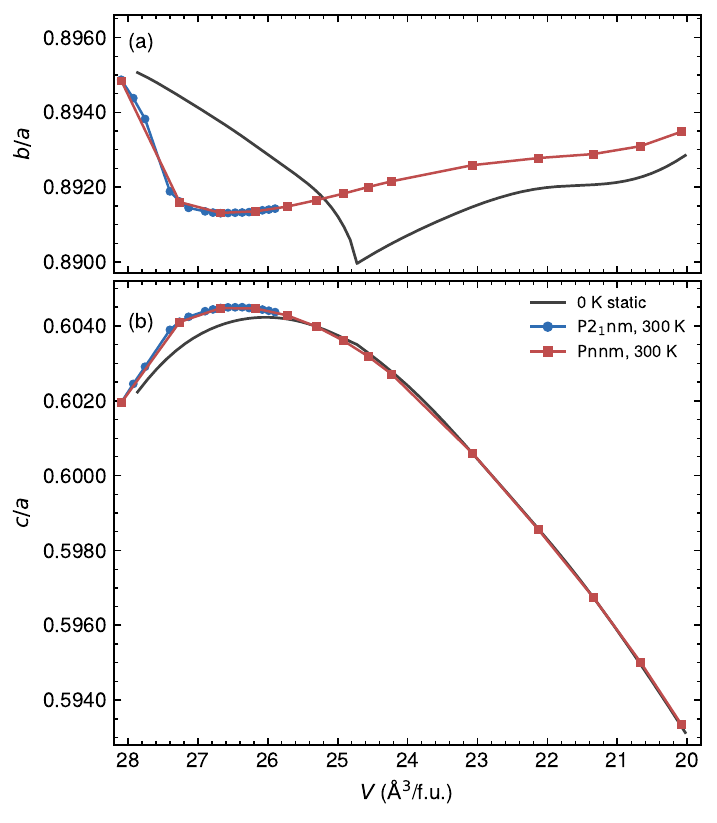}
    \caption{(a) $b/a$ and (b) $c/a$ cell-shape ratios of the 300~K average structures compared with the ordered 0~K static structures. The ratios are plotted against volume per $\delta$-AlOOH.}
    \label{fig:supp-cell-shape}
\end{figure}

\begin{figure}[htbp]
    \centering
    \includegraphics[width=.48\textwidth]{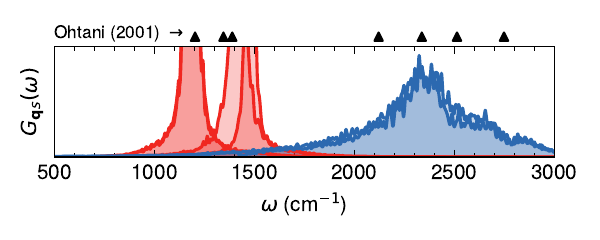}
    \caption{Power spectra, $G_{\mathbf{q}s}(\omega)$, of selected OH-stretching and bending bands at $\Gamma$ at 0~GPa and 300~K, compared with measured Raman peak positions under the same conditions.}
    \label{fig:supp-qproj}
\end{figure}

\begin{figure}[htbp]
    \centering
    \includegraphics[width=.62\textwidth]{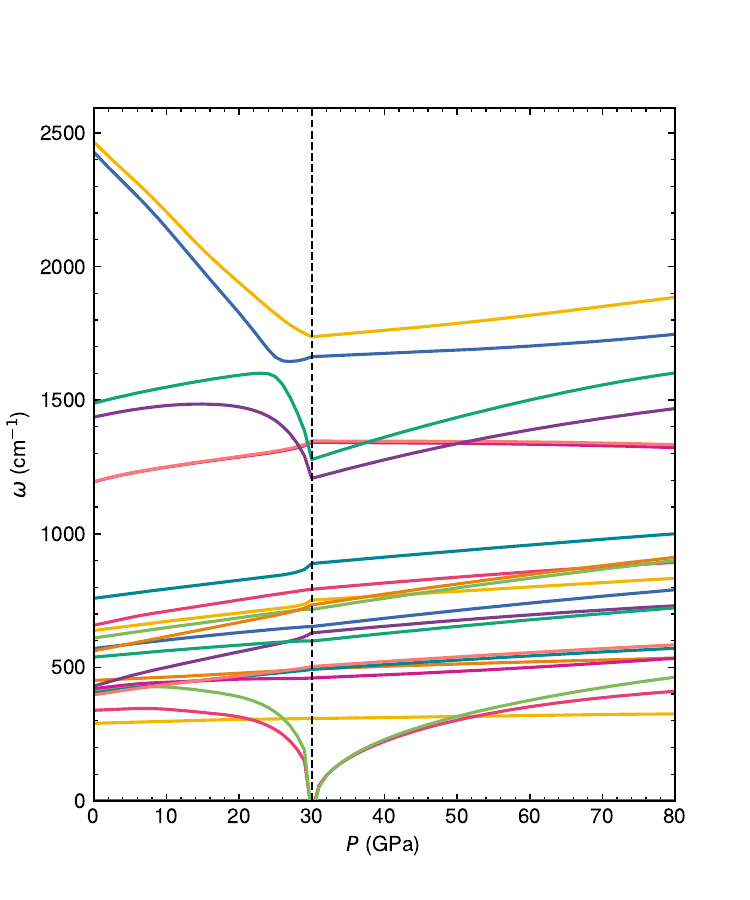}
    \caption{Pressure dependence of the ordered 0~K harmonic frequencies, $\omega_{\mathbf{q}s}$, at $\Gamma$ from static strongly constrained and appropriately normed (SCAN)-based Deep Potential (SCAN-DP) calculations. Each curve traces one branch matched continuously in pressure. The dashed vertical line marks the approximate static $\mathrm{P}2_1\mathrm{nm}$-to-$\mathrm{Pnnm}$ boundary at $P = 30$~GPa, where the ordered-zone-center frequencies soften toward the static symmetrization pressure.}
    \label{fig:supp-wavenumbers-0k}
\end{figure}

\clearpage

\section{Representative ambient-pressure quasiparticle data}

\begin{table}[htbp]
\centering
\caption{$\Gamma$-point quasiparticle frequency, $\tilde\omega_{\mathbf{q}s}$, frequency shift relative to the harmonic phonon, $\Delta \omega_{\mathbf{q}s} = \tilde\omega_{\mathbf{q}s} - \omega_{\mathbf{q}s}$, and linewidth, $\Gamma_{\mathbf{q}s}$, predicted by SCAN-based deep-potential molecular dynamics (SCAN-DPMD) at 0~GPa and 300~K for the $\mathrm{P}2_1\mathrm{nm}$ average structure.}
\label{tab:quasiparticle-ambient}
\begin{ruledtabular}
\begin{tabular}{lrrrl}
        & $\tilde\omega_{\mathbf{q}s}$ & $\Gamma_{\mathbf{q}s}$ & $\Delta\omega_{\mathbf{q}s}$ & irrep \\
   \#   & (cm$^{-1}$) & (cm$^{-1}$) & (cm$^{-1}$) \\
\hline
3  &      286 &        1 &          $   -3$ &    $\mathrm{A}_2$ \\
4  &      313 &       46 &          $  -24$ &    $\mathrm{A}_1$ \\
5  &      393 &        5 &          $    3$ &    $\mathrm{B}_1$ \\
6  &      400 &        6 &          $    2$ &    $\mathrm{A}_2$ \\
7  &      401 &       28 &          $   -2$ &    $\mathrm{B}_2$ \\
8  &      414 &        5 &          $    0$ &    $\mathrm{A}_1$ \\
9  &      422 &       14 &          $    2$ &    $\mathrm{A}_1$ \\
10 &      433 &       23 &          $  -16$ &    $\mathrm{B}_2$ \\
11 &      525 &        6 &          $   -8$ &    $\mathrm{B}_1$ \\
12 &      557 &        6 &          $    4$ &    $\mathrm{B}_2$ \\
13 &      558 &        8 &          $   -6$ &    $\mathrm{A}_2$ \\
14 &      597 &       13 &          $   -6$ &    $\mathrm{A}_1$ \\
15 &      621 &       13 &          $  -11$ &    $\mathrm{A}_1$ \\
16 &      650 &       16 &          $    1$ &    $\mathrm{B}_2$ \\
17 &      744 &       17 &          $   -8$ &    $\mathrm{B}_2$ \\
18 &     1185 &       37 &          $    2$ &    $\mathrm{A}_2$ \\
19 &     1185 &       40 &          $    1$ &    $\mathrm{B}_1$ \\
20 &     1416 &       35 &          $  -11$ &    $\mathrm{B}_2$ \\
21 &     1477 &       16 &          $   -0$ &    $\mathrm{A}_1$ \\
22 &     2329 &      428 &          $ -140$ &    $\mathrm{A}_1$ \\
23 &     2373 &      449 &          $ -132$ &    $\mathrm{B}_2$ \\
\end{tabular}
\end{ruledtabular}
\end{table}